\documentclass[10pt]{iopart}
\usepackage{graphicx}

\begin{document}

\title{Shape coexistence in the microscopically guided interacting boson model}

\author{K. Nomura$^1$\footnote{Present address: Physics Department,
Faculty of Science, University of Zagreb, HR-10000 Zagreb, Croatia},  
T. Otsuka$^{2,3,4}$, and P. Van Isacker$^1$}

\address{$^1$ Grand Acc\'el\'erateur National d'Ions Lourds, CEA/DSM-CNRS/IN2P3, B. P. 55027, F-14076 Caen Cedex 5, France}
\address{$^2$ Department of Physics, University
of Tokyo, Hongo, Bunkyo-ku, Tokyo 113-0033, Japan}
\address{$^3$ Center for Nuclear Study, University
of Tokyo, Hongo, Bunkyo-ku, Tokyo 113-0033, Japan}
\address{$^4$ National Superconducting Cyclotron Laboratory, Michigan
State University, East Lansing, Michigan 48824, USA}
\ead{nomura@ikp.uni-koeln.de}
\vspace{10pt}
\begin{indented}
\item[]November 2015
\end{indented}

\begin{abstract}
Shape coexistence has been a subject of great interest in nuclear
 physics for many decades. 
In the context of the nuclear shell model, intruder
 excitations may give rise
 to remarkably low-lying excited $0^+$ states associated with different intrinsic
 shapes. 
In heavy open-shell nuclei, the dimension of the shell-model
 configuration space that includes such intruder excitations becomes 
 exceedingly large, 
thus requiring a drastic truncation scheme. 
Such a framework has been provided by the interacting boson
 model (IBM). 
In this article we address the phenomenon of shape coexistence and its relevant
 spectroscopy from the point of view of the IBM. 
A special focus is placed on the method developed recently 
which makes use of the link between the IBM and the self-consistent
 mean-field approach based on the nuclear energy density functional. 
The method is extended to deal with various intruder configurations associated
 with different equilibrium shapes. 
We assess the predictive power of the method and suggest possible
 improvements and extensions, by considering illustrative examples
 in the neutron-deficient Pb region, where shape coexistence has been
 experimentally studied. 
\end{abstract}



\section{Introduction\label{sec:intro}}

The study of shape coexistence and related collective excitations in atomic
nuclei has been 
a theme of major interest in low-energy nuclear structure physics for
more than half a century \cite{morinaga56,heyde83,heyde87,wood92,andreyev00,heyde11}. 
In specific regions of the periodic table, unexpectedly low-lying 
excited $0^+$ states close in energy to the $0^+$ ground state have been
observed. 
In the spherical shell model \cite{heyde87,federman77,heyde85}, the emergence of 
the low-lying excited $0^+$ states has been often attributed to
multi-particle-multi-hole intruder excitations. 
This scenario may hold in mass regions near shell closure,
the example being the neutron-deficient lead region. 
In this case, two or four protons in the $Z=50-82$ major shell can be 
excited to the next major shell. 
The correlation between valence protons and neutrons is enhanced due to
this cross-shell excitation of protons, resulting in the lowering of the
excited $0^+$ configurations. 
This effect becomes prominent especially around the middle of the
neutron major shell $N=104$, where the number of valence neutrons is
maximal.

The shell-model approach \cite{otsuka01,caurier05} 
has been a reliable means of studying spectroscopic properties relevant to shape coexistence in a quantitative way. 
However, the dimension of the configuration space becomes 
exceedingly large for open-shell heavy-mass nuclei, requiring a drastic
truncation scheme to make the problem tractable. 
Such a framework has been provided by the interacting boson model (IBM)
\cite{IBM}, in which the correlated pairs of valence nucleons, 
$S$ ($J^{\pi}=0^+$) and $D$ ($2^{+}$), are mapped onto $s$ and $d$
bosons, respectively \cite{arima77,OAIT,OAI}.  
In the IBM, the number of proton (neutron) bosons, denoted by $N_{\pi}$ ($N_{\nu}$),
is equal to that of pairs of valence protons (neutrons) \cite{OAI}. 
The IBM is strongly connected to group theory: if the Hamiltonian
is expressed as a specific combination of 
Casimir operators, the corresponding system has 
a dynamical symmetry associated with a certain intrinsic structure and is
exactly solvable 
\cite{IBM}. 
In the simplest version of the IBM consisting of only $s$ and $d$
bosons, for instance, three possible dynamical
symmetries, U(5), SU(3) and SO(6), emerge, corresponding to vibrational, rotational and
$\gamma$-unstable states of the quadrupole mode. 
The IBM has been well established for systematic and straightforward, albeit purely
phenomenological, calculations of nuclear collective excitation
energies and electromagnetic transition rates in medium- and heavy-mass nuclei. 

A method to incorporate intruder excitations in the IBM was 
first presented by Duval and Barrett \cite{duval81,duval82}.  
They proposed to associate the different shell-model spaces of $0p-0h$, $2p-2h$,
$4p-4h$, $\ldots$ excitations with the corresponding boson spaces 
comprising 
$N$, $N+2$, $N+4$, $\ldots$ bosons, with $N$ being the total boson
number $N=N_{\pi}+N_{\nu}$, and the different boson subspaces are
subsequently mixed by a specific Hamiltonian (see Subsec.~\ref{sec:DB}). 
The Duval-Barrett procedure has
been applied to the study of spectroscopic properties mainly in the Pb region
\cite{fossion03,hellemans05,hellemans08,garciaramos14hg,garciaramos14}.  
%
The concept of intruder analog states was introduced by means of a new
quantum number, the $I$-spin or intruder spin  
\cite{heyde92,heyde94,coster96},
while geometrical features of configuration mixing IBM were addressed \cite{frank04,frank06,morales08}. 
Similar to previous IBM calculations, the drawback of the
configuration mixing IBM approach is that 
its Hamiltonian parameters are constrained only by known
experimental data on intruder states.

Let us stress that many attempts have been made to
establish a link between the IBM and more fundamental nuclear
structure models \cite{iachello87,klein91}. 
The prescription widely taken is the one developed by Otsuka 
et al. \cite{OAI}, often referred to as the Otsuka-Arima-Iachello (OAI)
mapping, where a seniority-based shell-model state, expressed by 
low-spin $S$, $D$, $\ldots$ pairs, is mapped onto the equivalent boson 
$s$, $d$, $\ldots$ states. 
This method is, however, rather restricted to nuclei with near-spherical 
or $\gamma$-soft shapes \cite{mizusaki96}, where the
simple seniority classification is a reasonable approximation. 
The microscopic derivation of the IBM for 
nuclei with general shapes, particularly strongly deformed ones, 
has been under 
active investigation in the 1980s, but since then 
without much progress. 

An entirely new method of deriving the IBM Hamiltonian was 
presented in Ref.~\cite{nom08}, whose authors proposed to derive the 
IBM Hamiltonian by mapping the total energy surface, calculated within
the self-consistent mean-field [e.g., Hartree-Fock-Bogoliubov (HFB)]
approach using a microscopic nuclear energy 
density functional (EDF) onto the equivalent bosonic total energy
surface. 
Self-consistent mean-field models on the basis of a given non-relativistic 
\cite{bender03,drut10,dobaczewski11,dobaczewski12} or relativistic
\cite{ring96,vretenar05,nik11} EDF are well established for the
description of matter and bulk ground-state properties of almost all nuclei, including
binding energies, charge radii {\it etc}. Their accuracy is remarkable 
and they can thus be suitable as microscopic input to the IBM for any
deformed configuration. 
In this way, the IBM has been shown to be derived only from nucleonic degrees of freedom in principle for any  
deformed state, including near spherical or weakly deformed
\cite{nom10}, strongly-deformed rotational \cite{nom11}, and
$\gamma$-soft \cite{nom12tri} regimes. 
Moreover, this allows one to have access to exotic isotopes \cite{albers12}. 
Recently, the method of Ref.~\cite{nom08} was extended to handle shape
coexistence \cite{nom12sc} and applied to the neutron-deficient Pb \cite{nom12sc} and
Hg \cite{nom13hg} nuclei and to some neutron-rich nuclei with mass $A\approx 100$ \cite{albers13,thomas13}.

We mention that calculations of nuclear spectroscopic properties have also been
extensively pursued within beyond mean-field approaches \cite{bender03,bender04,rayner04,yao13}, by
projecting the mean-field solution of an EDF calculation onto 
states with good symmetry and by taking into account
quantum fluctuations around the mean-field minimum
\cite{bender03,dobaczewski11}. 

This contribution gives 
a detailed description of the combined mean-field and IBM approach of
Ref.~\cite{nom08}, more in particular, its recent 
extension to the study of shape coexistence \cite{nom12sc}. It also addresses the
possibilities of the method for the future. 
The article is organized as follows: we give a brief overview of the
relevant IBM framework in Sec.~\ref{sec:IBM}, describe the theoretical
framework of the proposed method in Secs.~\ref{sec:theory} and \ref{sec:sc}
including some illustrative results for neutron-deficient Pb and Hg isotopes. 
Finally, Sec.~\ref{sec:summary} is devoted to concluding remarks and
possible improvements and extensions of the method.

\section{Interacting boson model for shape coexistence \label{sec:IBM}}

\subsection{Implementation of intruder configurations in the IBM\label{sec:DB}}

\begin{figure}[ctb!]
\begin{center}
\includegraphics[width=12.0cm]{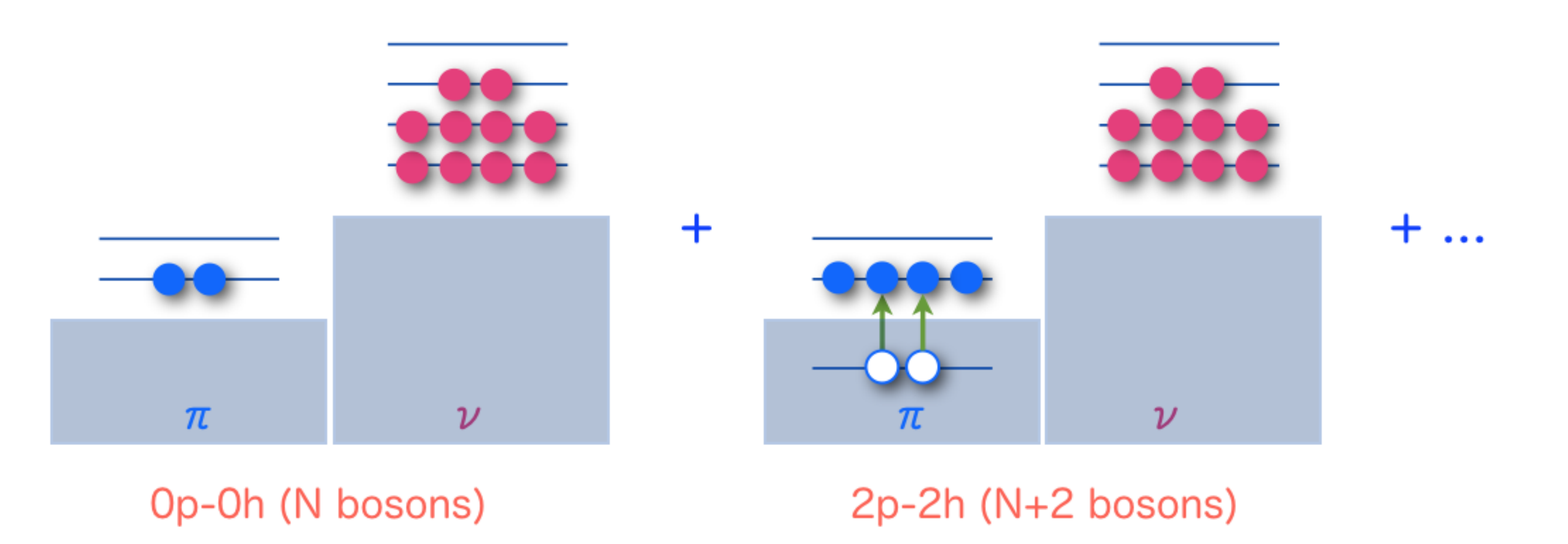}
\end{center}
\caption{(Color online) Illustration of the configuration mixing in the
 IBM framework.}
\label{fig:sc}
\end{figure}

We first give a brief overview of the Duval-Barrett technique
\cite{duval81,duval82} which allows one to mix different IBM configurations
to describe low-lying excited intruder states. 
Throughout this article, we assume only proton excitations, because in
all the cases considered here the proton number is close to the shell  
closure and cross-shell excitations of the protons can therefore be
assumed more probable than those of neutrons. 

In counting the number of bosons, Duval and Barrett followed the 
assumption that particle-like and hole-like
bosons are not distinguished \cite{duval81,duval82}. 
Under this assumption, as the excitation of one proton pair (boson) increases the boson
number by 2, the $np-nh$ ($n=0,2,4,\ldots$) configurations differ
from each other in boson number by 2. 
Hence, to describe a system consisting of different intruder excitations, the boson 
Hilbert space is extended to the space 
$[N_{\pi}\otimes N_{\nu}]\oplus[(N_{\pi}+2)\otimes
N_{\nu}]\oplus[(N_{\pi}+4)\otimes N_{\nu}]\oplus\cdots$, a direct sum of
the orthogonal subspaces of the $(N_{\pi},N_{\nu})$,
$(N_{\pi}+2,N_{\nu})$, 
$(N_{\pi}+4,N_{\nu})$, $\ldots$ boson systems,
respectively (see Fig.~\ref{fig:sc}). 
In the following, the subspace $[(N_{\pi}+n)\otimes N_{\nu}]$
($n=0,2,4,\ldots$) is denoted simply as $[N+n]$ with
$N=N_{\pi}+N_{\nu}$.

Let us consider the  $^{186}$Pb nucleus as an example, as it is empirically known for the
coexistence of three intrinsic shapes: spherical, oblate and prolate, corresponding to $0p-0h$, $2p-2h$ and
$4p-4h$ shell-model configurations, respectively \cite{andreyev00}. 
The total Hamiltonian is expressed as: 
\begin{eqnarray}
\label{eq:ham-cm}
\hat H=\hat P_0\hat H_0\hat P_0+\hat P_{2}(\hat
H_{2}+\Delta_{2})\hat P_{2}+\hat P_{4}(\hat H_{4}+\Delta_{4})\hat
P_{4}+\hat V^{\rm mix}_{0,2}+\hat V^{\rm mix}_{2,4}, 
\end{eqnarray}
where $\hat H_{n}$ ($n=0,2,4$) represent the unperturbed Hamiltonians for the
proton $np-nh$ excitations in the shell model, and 
$\hat V^{\rm mix}_{n,n+2}$ ($n=0,2$) stand for the interaction
terms mixing the $[N+n]$ and $[N+n+2]$ subspaces, 
defined as 
\begin{eqnarray}
\label{eq:mix}
 \hat V^{\rm
 mix}_{n,n+2}=\hat
 P_{n+2}(\omega^s_{n,n+2}s^{\dagger}_{\pi}s^{\dagger}_{\pi}+\omega^d_{n,n+2}d^{\dagger}_{\pi}d^{\dagger}_{\pi})^{(0)}\hat
 P_{n}+h.c., 
\end{eqnarray}
where $\omega^s_{n,n+2}$ and $\omega^d_{n,n+2}$ represent the strengths of mixing. 
A direct coupling between $0p-0h$ and $4p-4h$ spaces is absent with
two-body nuclear forces and is not considered here. 
The operators $\hat P_{n}$ ($n=0,2,4$) are projectors onto the 
subspace $[N+n]$, and $\Delta_{n+2}$ ($n=0,2$) represent the energies needed to
excite one or two proton pair (or boson) across the shell closure $Z=82$. 
The parameters for each unperturbed Hamiltonian are fitted to the
corresponding intruder bands. 
The values of the Hamiltonian parameters can be different from one configuration
to another. 
The off-set energies $\Delta_{n+2}$ are usually determined to reproduce the bandhead
$0^+$ energies of intruder bands. 
The mixing strengths are often introduced
perturbatively with small values relative to the Hamiltonian parameters.

Initially, the Duval-Barrett procedure was applied to Hg isotopes
in Refs.~\cite{duval81,duval82}, followed by Refs.~\cite{barfield83,barfield84} for
the description of new data. 
It was also applied to different mass
regions, including the Mo \cite{duval82,sambataro82} and Cd \cite{sambataro82cd} isotopes. 
An application to Ge isotopes was presented
\cite{duval83} with
separate treatment of particle-like and hole-like bosons in the mixing
interaction. 
More recently, both spectroscopic and ground-state properties of 
nuclei showing manifest shape coexistence 
were extensively studied for the chains of Pb
\cite{fossion03,hellemans05,hellemans08}, Hg \cite{garciaramos14hg} and
Pt \cite{garciaramos14} isotopes. 
On the other hand, the evidence for coexistence is somewhat hindered in
Pt isotopes, and whether or not the
intruder configuration is necessary in the IBM framework for Pt
isotopes is still a matter of controversy \cite{garciaramos14,mccutchan06,garciaramos09,garciaramos11,nom11pt}. 

\subsection{Microscopy of configuration-mixing IBM from a shell-model perspective}

Since the Duval-Barrett method has been used rather phenomenologically, its microscopic
foundation has also been addressed. 
Van Isacker et al. \cite{isacker86} derived the strengths of the mixing interaction
and the off-set energy for Cd and Hg isotopes from a simple shell-model interaction by using
the OAI mapping with the generalized seniority scheme, and pointed out
that, generally, while the derived off-set energy was consistent with the one
obtained from a phenomenological fit to data, the derived mixing strengths
were larger than the phenomenological ones. 
In Refs.~\cite{heyde87,wood92} a semi-empirical expression for  
the energy of the intruder $0^+$ state was proposed, composed
of the unperturbed separation of the regular and 
intruder configurations (off-set energy), of the monopole 
correction to the proton single-particle energy, and of the energy gained by
the pairing and the proton-neutron quadrupole correlations. 

\subsection{Group-theoretical aspects of shape coexistence}

As the IBM is closely linked to group theory, a symmetry-dictated
analysis of shape coexistence is also possible. 
The concept of intruder analog 
states, labelled by the $I$-spin quantum number, was introduced by Heyde and 
collaborators \cite{heyde92,heyde94} as a new class of  classification
scheme relevant to 
particle-hole excitations. 
A particle-like boson is represented by a spin $I=1/2$ and its projection
$I_z=+1/2$ and a hole-like boson by a spin $I=1/2$ and its projection
$I_z=-1/2$. 
The algebra for $I$-spin is analogous to that for isospin for fermions
and the one for the $F$-spin in the proton-neutron 
interacting boson model (IBM-2) \cite{OAI}.

The $I$-spin symmetry was used as a guide to constrain parameters of the
boson Hamiltonian in describing the structure of Pb isotopes
\cite{fossion03} where the Hamiltonians for the $2p-2h$ and $4p-4h$ 
configurations were fixed by taking the parameters from adjacent
even-even $Z=78$ (Pt) and $Z=74$ (W) isotopes, already known to give a good
description of spectroscopic data. 
The same idea was applied to the Po isotopes 
\cite{decoster99}.

In addition to the $I$-spin formalism, the algebraic features of 
particle-hole excitation modes relevant to shape coexistence were 
further investigated in the context of the couplings of different
dynamical symmetries: U(5)-SU(3) \cite{decoster99}, U(5)-SO(6) \cite{lehmann97},
and SO(6)-SU(3) \cite{decoster97}.

\subsection{Geometry and phases in the IBM with configuration mixing}

A deformation energy surface analogous to the one in the collective
model can be defined in the IBM, allowing one to {\it visualize} the
corresponding geometry and phase transitions \cite{gilmore78,dieperink80}. 
A boson coherent state, or intrinsic 
state of the bosonic system $|\phi\rangle$, was introduced in 
Refs.~\cite{dieperink80,ginocchio80,bohr80}, expressed, up to
a normalization factor, as: 
\begin{eqnarray}
\label{eq:coherent}
 |\phi\rangle=\frac{1}{\sqrt{N!}}(\lambda^{\dagger})^{N}|{\textnormal{o}}\rangle, \quad{\rm
 with}\quad
\lambda^{\dagger}=s^{\dagger}+\sum_{\mu}\alpha_{\mu}d_{\mu}^{\dagger}, 
\end{eqnarray}
where proton and neutron bosons are not distinguished for simplicity,
and 
where the five coefficients $\alpha_{\mu}$ are related to the orientation
of the nucleus (three Euler angles) and two intrinsic deformation
variables $\beta$ and $\gamma$, equivalent to
those in the collective model \cite{bohr75}. 
More explicitly, $\alpha_{\pm 2}=2^{-1/2}\beta\sin{\gamma}$, $\alpha_{\pm 1}=0$ and
$\alpha_{0}=\beta\cos{\gamma}$, where $|{\textnormal{o}}\rangle$ represents the boson vacuum (or inert core). 
The expectation value of a given IBM Hamiltonian $\hat H_{b}$ in the coherent
state, if written as a function of deformation parameters $|\phi(\beta,\gamma)\rangle$, provides one with the bosonic total energy surface
$\langle\phi(\beta,\gamma)|\hat H_b|\phi(\beta,\gamma)\rangle$. 
Equivalent to the collective model, the bosonic energy surface generates 
all possible quadrupole shapes: spherical (if $\beta=0$), prolate (if $\beta\neq 0$ and $\gamma=0^{\circ}$), oblate
(if $\beta\neq 0$ and $\gamma=60^{\circ}$) and triaxial (otherwise) shapes.

Frank et al. \cite{frank04} studied the geometry of the
configuration-mixing IBM, with energy surfaces with more than one minimum. 
In Ref.~\cite{frank04} the intrinsic state for $np-nh$ 
($n=0,2,4,\ldots$) excitations was defined in the model space 
$[N]\oplus[N+2]\oplus[N+4]\oplus\cdots$.  
For the system of three configurations, for instance, the expectation
value of the configuration-mixing IBM Hamiltonian in Eq.~(\ref{eq:ham-cm}) is
expressed as the following $3\times 3$ matrix \cite{frank04}: 
\begin{eqnarray}
\label{eq:pes}
  {\cal E}(\beta,\gamma)=\left(
\begin{array}{ccc}
E_0(\beta,\gamma) & \Omega_{0,2}(\beta,\gamma) & 0 \\
\Omega_{2,0}(\beta,\gamma) & E_{2}(\beta,\gamma)+\Delta_{2} & \Omega_{2,4}(\beta,\gamma)
 \\
0 & \Omega_{4,2}(\beta,\gamma) & E_{4}(\beta,\gamma)+\Delta_{4} \\
\end{array}
\right), 
\end{eqnarray}
where the diagonal and off-diagonal matrix elements  stand for
the expectation values of the unperturbed Hamiltonians and 
the mixing interactions, respectively. 
The three eigenvalues of ${\cal E}(\beta,\gamma)$ lead to 
specific energy surfaces, which are different from each other in
topology depending mainly on the Hamiltonian
parameters and off-set energies \cite{morales08}. 
The lowest eigenvalue at each set 
of $\beta$ and $\gamma$ deformations is taken as the
relevant energy surface \cite{frank04}. 
%
%
It displays three minima, at spherical,
prolate and oblate deformation, if the parameters fitted to the spectroscopic
properties of the neutron-deficient Pb are taken in the Hamiltonian of
Eq.~(\ref{eq:ham-cm}). 
Along the same line, quantum phase transitions of the configuration-mixing IBM
were analyzed \cite{frank06}, and tested in some Pt nuclei \cite{morales08}.

\section{Mean-field derivation of the IBM \label{sec:theory}}

We now turn to the description of the HFB-to-IBM mapping method \cite{nom08}. 
We start with a brief note on the mean-field calculation in
Subsec.~\ref{sec:mf}, then show how the mapping is carried out
(Subsecs.~\ref{sec:sg}, \ref{sec:rot} and \ref{sec:triaxial}). 
We demonstrate that the method is able to describe reasonably well the low-energy
spectroscopy of  
Sm and Ba isotopes in Subsec.~\ref{sec:proof}, representative of
axially-deformed and $\gamma$-soft nuclei, respectively.

\subsection{Mean-field calculation\label{sec:mf}}

\begin{figure}[ctb!]
\begin{center}
  \includegraphics[width=8cm]{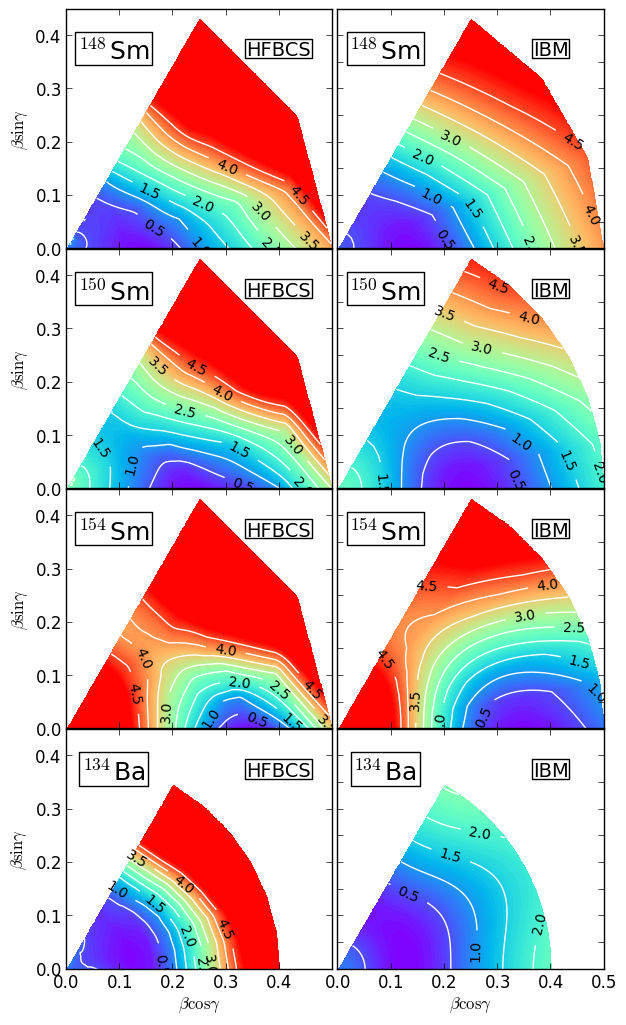}
  \caption{(Color online) Total 
  ($\beta$, $\gamma$) energy surfaces of the $^{148,150,154}$Sm and $^{134}$Ba isotopes. On the
  left-hand side (HFBCS) are microscopic energy surfaces with the Skyrme
  SkM* interaction, while on the right-hand side are the mapped IBM energy
  surfaces. The contour lines join
 points with the same energy (in MeV) and the color scale varies in
 steps of 50 keV. The energy difference between neighboring contours is
 500 keV. Data points are from Ref.~\cite{nom10}.}
  \label{fig:sm-pes}
\end{center}
\end{figure}

The first step is a standard constrained self-consistent
mean-field calculation, e.g., 
in the Hartree-Fock-Bogoliubov (HFB) framework,  based on a relativistic or
non-relativistic EDF that is already shown to be valid in the
global description of nuclear structure phenomena (see, e.g., reviews in 
Refs.~\cite{bender03,nik11}). 
In this case, the constraints are on mass
quadrupole moments related to the axial deformation $\beta$ and
non-axial deformation $\gamma$ \cite{bohr75}. 
For a given set of collective coordinates $(\beta,\gamma)$, the HFB
calculations are performed to obtain an energy surface, which refers to
the total mean-field energy without any
symmetry projection. 
Neither the mass parameter nor the collective potential is considered
explicitly like in some studies deriving a collective Hamiltonian from a set of EDF mean-field 
calculations \cite{nik11,delaroche10}.

Examples of microscopic total energy surfaces are shown 
on the left-hand side of 
Fig.~\ref{fig:sm-pes} for 
$^{148,150,154}$Sm and $^{134}$Ba nuclei. 
In these particular cases, the calculation was made with the
constrained Hartree-Fock plus BCS (HFBCS) method \cite{ev8} using the Skyrme \cite{skyrme,vautherin72}
EDF with the SkM* parametrization \cite{skms}. 
One sees in the figure that the HFBCS calculation gives a energy
minimum near the origin for $^{148}$Sm, a deep minimum 
at $\beta\approx 0.35$ for $^{154}$Sm, and for $^{150}$Sm a minimum soft
in $\beta$ direction. 
For $^{134}$Ba, the HFBCS calculation predicts a $\gamma$-soft structure. 

\subsection{Mapping onto the boson space \label{sec:sg}}


Having the HFB total energy surface 
$\langle\phi_{f}(\beta,\gamma)|\hat H_f|\phi_{f}(\beta,\gamma)\rangle$ for each nucleus,
with $|\phi_{f}(\beta,\gamma)\rangle$ representing the HFB solution for a given 
deformation $(\beta,\gamma)$, we subsequently map this energy surface to the corresponding
energy surface of the IBM. 
To be more specific, the IBM Hamiltonian is constructed, i.e., the
IBM strengths parameters are fixed, so as to reproduce the HFB energy surface
at each set of ($\beta,\gamma$) as close as possible. 

The coherent state (\ref{eq:coherent}) in the IBM-2, denoted as
$|\phi_b\rangle$, is 
written as the product of proton $\lambda_\pi$ 
and neutron $\lambda_\nu$
boson condensates
$|\phi_b\rangle=(N_{\pi}!N_{\nu}!)^{-1/2}(\lambda_{\pi})^{N_{\pi}}(\lambda_{\nu})^{N_{\nu}}|{\rm
o}\rangle$
which invokes deformation variables for both protons ($\beta_{\pi},\gamma_{\pi}$) and
neutrons ($\beta_{\nu},\gamma_{\nu}$). 
We assume that the proton and neutron systems have the same
intrinsic shapes: 
$\beta_{\pi}=\beta_{\nu}\equiv\beta_b$ and
$\gamma_{\pi}=\gamma_{\nu}\equiv\gamma_b$. 
The four deformation parameters 
($\beta_{\pi},\beta_{\nu},\gamma_{\pi},\gamma_{\nu}$) could in principle
vary independently but, in 
realistic cases, introducing all of these variables 
entails too much complexity. 
Therefore, we assume equal deformations of the proton and neutron
systems, as is usual in the collective model \cite{bohr75}. 
The variable $\beta_b$ in the boson system is proportional to $\beta$
deformation in the collective model $\beta\propto\beta_b$ \cite{ginocchio80}, 
while $\gamma_b$ and $\gamma$ have the same meaning. 

Under these conditions, 
we equate the HFB and the IBM total energy surfaces: 
\begin{eqnarray}
\label{eq:hmap}
 \langle\phi_f(\beta,\gamma)|\hat H_f|\phi_f(\beta,\gamma)\rangle\sim
  \langle\phi_b(\beta_b,\gamma_b)|\hat H_b|\phi_b(\beta_b,\gamma_b)\rangle, 
\end{eqnarray}
with 
\begin{eqnarray}
\label{eq:bgmap}
 \beta=\beta_b/C_{\beta},\quad{\textnormal{and}}\quad \gamma=\gamma_b, 
\end{eqnarray}
where $C_{\beta}$ is a coefficient. 
Note that Eq.~(\ref{eq:hmap}) represents an approximate equality as it
is fulfilled within a limited range of ($\beta,\gamma$) plane, i.e.,
around the global minimum. 
We restrict ourselves to this range because it is most relevant to the low-energy
collective states. 
One should not try to reproduce the region far from the energy
minimum as the topology of the
microscopic energy surface around the region is determined by single-nucleon 
configurations that are, by construction, outside the model space of the
IBM consisting of only collective pairs of valence nucleons. 
For this reason, the IBM energy surface is, as shown later, always flat in the
region far from the minimum. 
The topology of the HFB total energy surface around global minimum should reflect essential
features of fermion many-body systems, such as the Pauli principle, the antisymmetrization, and
the underlying inter-nucleon interactions {\it etc}. 
Through the mapping procedure, these effects are supposed to be simulated by the IBM.

The IBM Hamiltonian that embodies the essentials of the underlying
fermionic interactions is taken to be of the form \cite{IBM,OAI}: 
\begin{eqnarray}
\label{eq:ham-sg}
 \hat H_b=\epsilon\hat n_d+\kappa\hat Q_{\pi}\cdot\hat Q_{\nu}, 
\end{eqnarray}
where the first and second  term represents, respectively, the  $d$-boson number operator ($\hat n_d=\hat n_{d\pi}+\hat
  n_{d\nu}$), with $\epsilon$ the single $d$-boson energy relative to the $s$-boson one, and the 
quadrupole-quadrupole interaction with 
strength $\kappa$. 
The quadrupole operator is $\hat Q_{\tau}=d^{\dagger}_{\tau}s_{\tau}+s^{\dagger}_{\tau}\tilde
d_{\tau}+\chi_{\tau}(d^{\dagger}_{\tau}\tilde d_{\tau})^{(2)}$
($\tau=\pi$ or $\nu$), with $\chi_{\tau}$ a parameter. 
The free parameters are 
$\epsilon$, $\kappa$, 
$\chi_{\pi}$, and $\chi_{\nu}$, 
plus the coefficient $C_{\beta}$ (see Eq.~(\ref{eq:bgmap})), and are to be fitted to the HFB total energy surface.

A technical detail in the mapping is that the IBM
parameters are determined 
unambiguously by exploiting the technique of Wavelet Transform
\cite{wavelet}, which has been developed in the field of signal processing. 
More specifically, the Wavelet Transform of an IBM energy surface in the
relevant range of ($\beta,\gamma$) 
is fitted to the corresponding one of an HFB energy surface by the simplex method. 
We note that a straightforward use of $\chi$-square fit does not work: 
Due to the local 
topology of the microscopic energy surface that is irrelevant for 
low-energy collective excitations, the $\chi$-square fit can result in many 
different sets of IBM parameters that do not have 
physical sense. 
The Wavelet Transform, in contrast, has the advantage of extracting the most
relevant information of a given signal \cite{wavelet}, which is, in the
present case, the topology of the HFB energy surface around its minimum. 

On the right-hand side of 
Fig.~\ref{fig:sm-pes}, we observe that the mapped IBM energy surfaces of
all the nuclei considered 
reproduce the topology of the HFBCS counterparts in the vicinity of the absolute minima. 
We confirm that, compared to the original HFBCS energy surface, the one
obtained from a mapped IBM looks generally flat 
in the region far from the minimum. 
As discussed before, this is a consequence of the limited model space and/or the finite boson
number in the IBM framework.

\subsection{Incorporating rotational response\label{sec:rot}}

In a strongly deformed nucleus, the rotational moment of inertia turns out to be systematically
underestimated in the present method \cite{nom08,nom10}. 
This partly stems from the fact
that the response of nucleonic intrinsic state to rotation 
is significantly different from that of the corresponding
bosonic intrinsic state \cite{nom11}. 
In order that the rotational response of fermionic 
and bosonic systems becomes similar, 
the term $\hat L\cdot\hat L$ was introduced in the IBM \cite{nom11},
where $\hat L$ stands for the angular momentum operator, and
was shown to play a crucial role to improve the description of
rotational spectrum in the strongly deformed nucleus \cite{nom11}. 
Thus, for a complete description of
quadrupole mode, the Hamiltonian in Eq.~(\ref{eq:ham-sg}) should include
this term. 
The $\hat L\cdot\hat L$ term affects the moment of inertia without changing other parameters
already fixed through the energy-surface mapping.  
In Ref.~\cite{nom11} it was proposed to derive the $\hat L\cdot\hat L$ strength
so that the cranking moment of inertia for the boson 
intrinsic state \cite{schaaser86}, calculated at the energy minimum, 
equals the cranking moment of inertia, e.g., of Thouless-Valatin
\cite{thouless62} type, in the HFB
framework at its corresponding minimum in the HFB total energy surface.

\subsection{
Additional degrees of freedom
\label{sec:triaxial}
}

Let us mention the possibility of introducing higher-order contributions in the
IBM Hamiltonian. 
This becomes relevant especially when the energy surface has a more
complex topology, e.g., $\gamma$-soft with a shallow non-axial minimum. 
To describe the $\gamma$-soft triaxiality in the IBM, it was shown
\cite{nom12tri} to be necessary to consider an interaction among up to
three bosons, i.e., a three-boson interaction. 
The three-boson interaction was already introduced in a phenomenological way
\cite{isacker81}, but was considered in Ref.~\cite{nom12tri} for the first time
from a microscopic perspective. 
An illustrative example is presented for $^{134}$Ba in
Fig.~\ref{fig:sm-pes}, where both the mapped IBM, which includes specific
three-body boson terms, and the HFBCS energy surfaces show
an almost totally $\gamma$-flat structure. 
One of the long-standing questions on nuclear shape has been 
whether a nucleus is $\gamma$ unstable
\cite{wilets56} or rather behaves as a rigid triaxial rotor
\cite{davydov58}. 
On the basis of the non-relativistic and 
relativistic EDF calculations of the total energy surface, it was found
\cite{nom12tri} that neither of the above two
geometrical pictures is realized but nuclei are somewhere in between. 
This finding is independent of the particular choice of the
EDF, and suggests an optimal IBM description of
triaxial shapes. 

Another case of interest arises if 
the reflection symmetry of the intrinsic shape of the nucleus is 
broken and negative-parity states emerge. In the IBM, in addition to the
positive-parity $s$ and $d$ bosons, it is
required to include negative-parity ($p,f,\ldots$) bosons. 
The octupole collective dynamics in a large number of rare-earth and 
light-actinide nuclei was addressed \cite{nom13oct} based on an axially-symmetric relativistic EDF 
calculation of the quadrupole-octupole deformation energy surface mapped
onto the $sdf$ IBM \cite{nom13oct,nom14}. 
This represents the first EDF-based calculation of shape phase
transitions, involving both quadrupole and octupole degrees of freedom. 

\subsection{Illustrative calculations \label{sec:proof}}

\begin{figure}[ctb!]
\begin{center}
  \includegraphics[width=10cm]{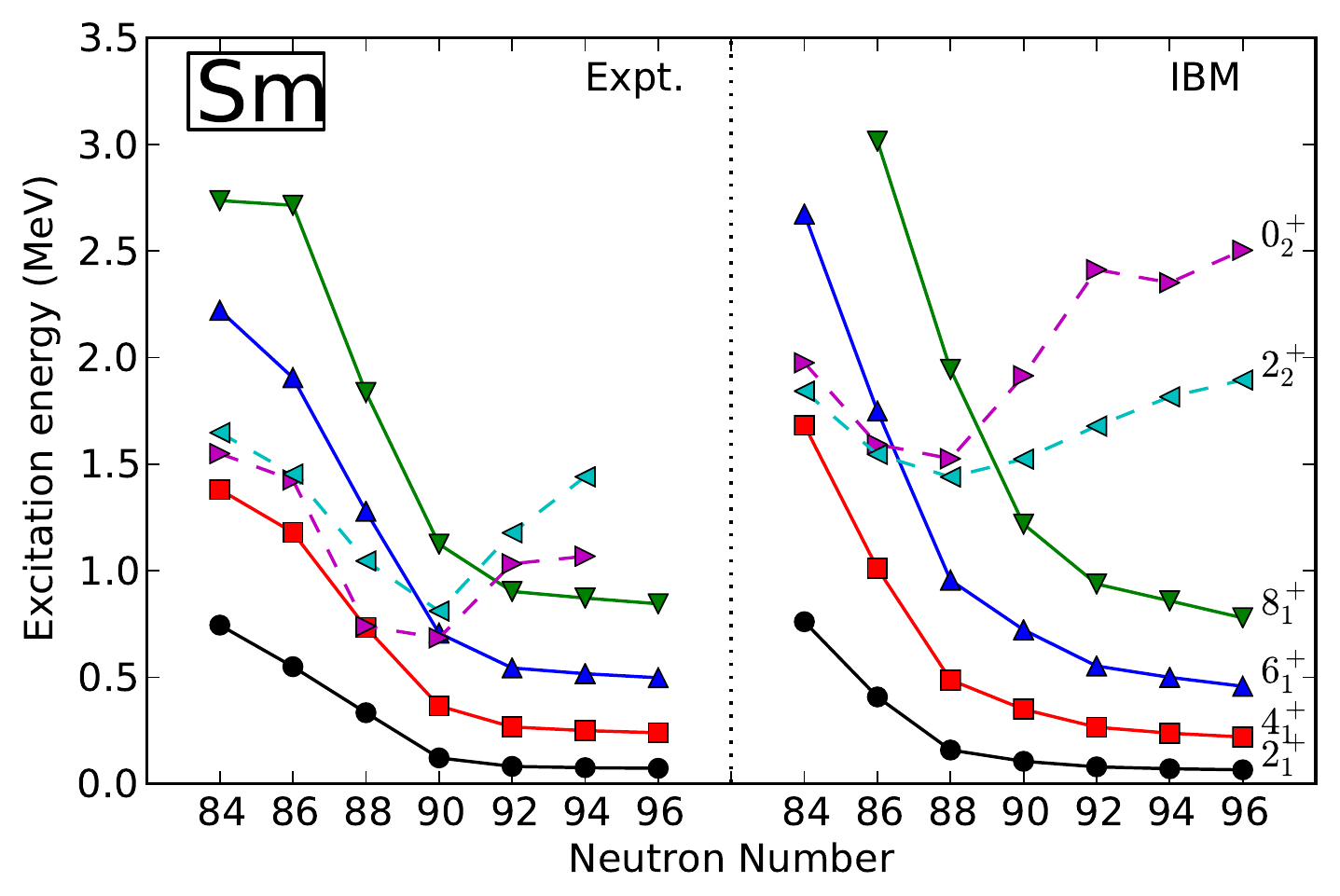}
  \caption{(Color online) Systematics of low-lying energy levels in the 
 $^{146-158}$Sm isotopes. Calculated and experimental energies are from
 Refs.~\cite{nom10,nom11} and from Ref.~\cite{data}, respectively. }
  \label{fig:sm-level}
\end{center}
\end{figure}

\begin{figure}[ctb!]
\begin{center}
  \includegraphics[width=10cm]{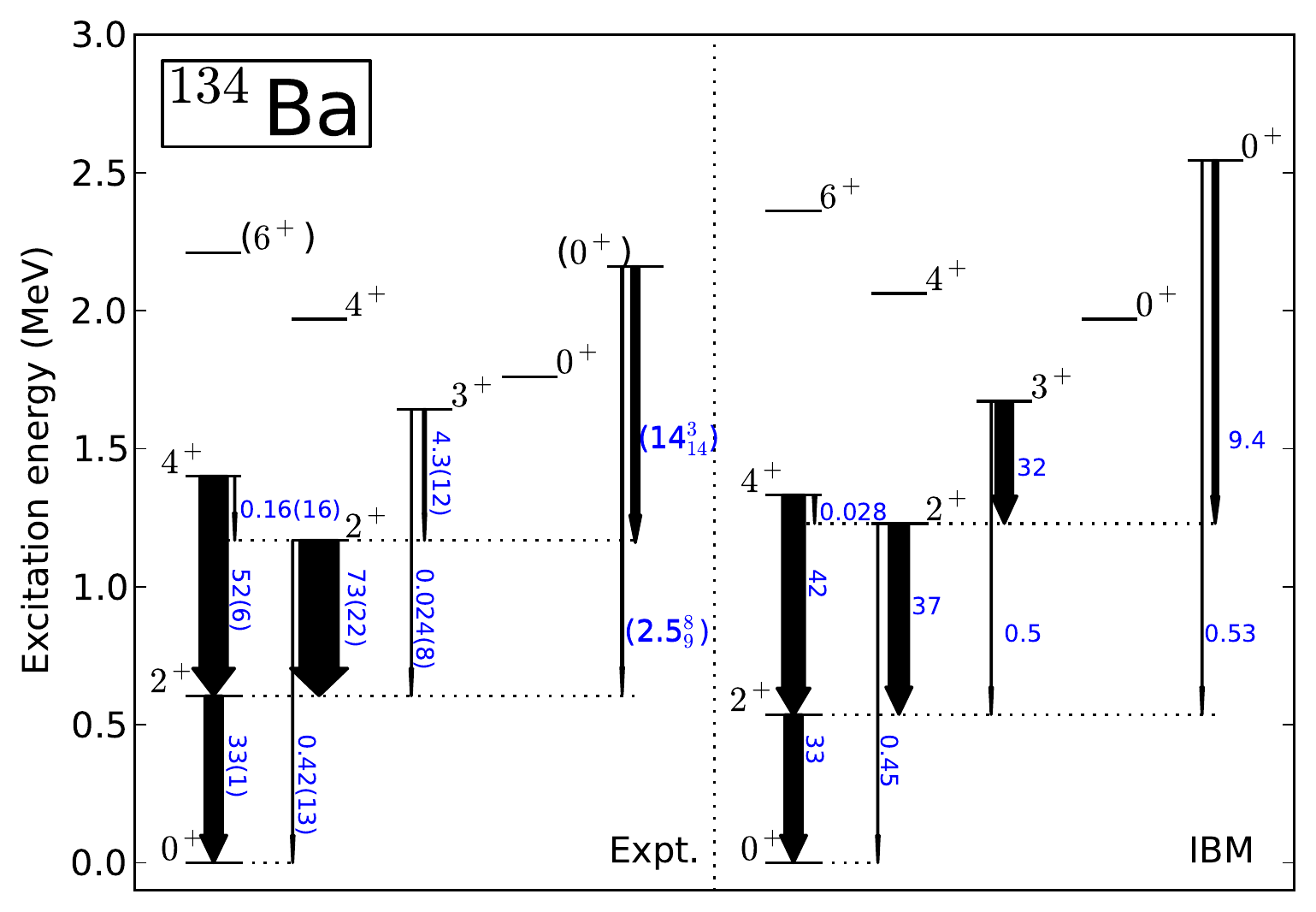}
  \caption{(Color online) Partial level scheme of  $^{134}$Ba, including $B$(E2) transition rates (in Weisskopf
 units). Experimental data have been taken from Refs.~\cite{data,pascu10}. As the 
 experimental $0^+_3$ state is uncertain \cite{data}, the corresponding $B$(E2)
 transition strengths are put in parentheses. The boson effective
 charges are fitted to the experimental $B$(E2$;2^+_1\rightarrow 0^+_1$) value.}
  \label{fig:ba-level}
\end{center}
\end{figure}

The diagonalization of the resultant IBM Hamiltonian, whose interaction strengths have been 
determined by the aforementioned ways, provides energies of and 
electromagnetic transition rates between excited states.

We show in Fig.~\ref{fig:sm-level} the calculated level energies of
some positive-parity yrast states of even-even $^{146-158}$Sm isotopes
and the
corresponding experimental data \cite{data}. 
In order to follow the structural evolution with increasing mass number
in a simple way, the following discussion is focused on the systematics of
yrast states up to spin $8^+$ and of two
non-yrast $0^+_2$ and $2^+_2$ states. 
From Fig.~\ref{fig:sm-level}, our calculation predicts that, for
$^{148}$Sm, the ratio $R_{4/2}=E(4^+_1)/E(2^+_1)$ is close to 2, 
being characteristic of vibrational nuclei, and that the yrast levels become
compressed with increasing $N$, resulting in a typical rotational band
with $R_{4/2}\approx 3.33$, especially for nuclei with $N\geq 90$. 
Generally, the calculation reasonably follows the experimental systematics. 
There is an overestimation of the non-yrast energy levels which is not
consistent with the present method and needs further
investigation.

We add a remark that it may not make much sense to discuss 
higher-spin ($J>8^+$) and/or higher-energy ($>$3 MeV) states in
the present framework. 
This is because the 
model space of the present IBM calculation comprising only $s$ and $d$
bosons may be rather 
limited to describe these states and because the IBM Hamiltonian is 
constructed only around the global minimum of a mean-field energy 
surface up to a few MeVs excitation energy. 
Furthermore, although a specific rotational cranking with non-zero
angular frequency is 
included to fix the strength of
the rotational correction term (see Subsec.~\ref{sec:rot}), we basically refer to a static
mean-field energy surface and it does not guarantee the validity of the model
description of higher-spin states with typically $J>8^+$. 

We show in Fig.~\ref{fig:ba-level} the partial level scheme of
$^{134}$Ba up to around 2.0-2.5 MeV, not considering the two additional $2^+$ levels at around 2 MeV
\cite{data,pascu10} as they are often discussed as possible
mixed-symmetry states \cite{pietralla08}, being outside the scope
of the present paper. 
In order to show that the 
present method is capable of describing basic features of a
$\gamma$-soft nucleus, we consider levels up to around 2.0-2.5 MeV, where the IBM
framework is generally valid \cite{IBM,mizusaki96}. 
From Fig.~\ref{fig:ba-level}, the overall spectroscopic information
provided by the present calculation, including 
the $0^+_{2,3}$ excitation
energies, the low-lying $2^+_2$ excitation energy and its strong E2 transition to
the $2^+_1$ relative to the $B$(E2;$2^+_1\rightarrow 0^+_1$) value 
are consistent with the interpretation of a $\gamma$-soft nucleus 
or SO(6) limit \cite{IBM} of the IBM.

At this point, we come to the conclusion that the HFB-to-IBM mapping
method is able to produce energy spectra that are typical for
transitional nuclei when moving away from closed shells into regions
with many valence protons and neutrons 
-- spherical vibrational, strongly deformed and
$\gamma$-soft shapes -- or, equivalently, three dynamical symmetries of the IBM. 
The derived IBM parameters have been shown \cite{nom08,nom10,nom11pt} to
be compatible with those used in previous 
phenomenological IBM calculations and those derived by the OAI mapping.


\section{Extension to shape coexistence \label{sec:sc}}

\subsection{Benchmark calculation: $^{186}$Pb\label{sec:bench}}

\begin{figure}[ctb!]
\begin{center}
\includegraphics[width=10.0cm]{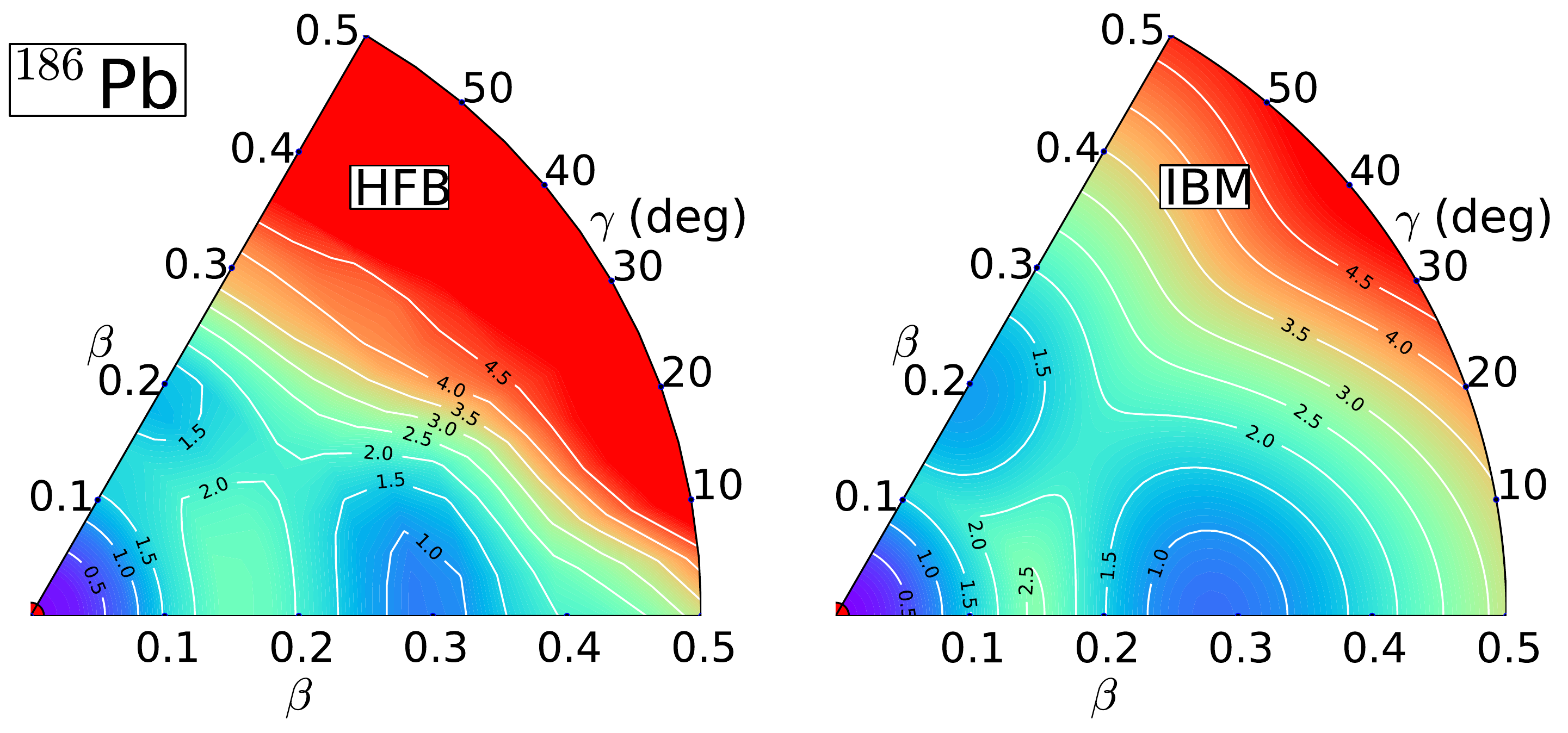}
\end{center}
\caption{(Color online) Same as for the caption to
 Fig.~\ref{fig:sm-pes}, but for the $^{186}$Pb nucleus. The results are
 taken from Ref.~\cite{nom12sc} and the HFB calculation is with the
 Gogny-D1M \cite{d1m} force. 
}
\label{fig:pb186-pes}
\end{figure}

In this section we demonstrate how the method can be extended to shape coexistence. 
Let us take the $^{186}$Pb nucleus as an illustrative example. 
On the left-hand side of Fig.~\ref{fig:pb186-pes} is shown the
($\beta,\gamma$) total energy surface obtained from the constrained
HFB method (see Ref.~\cite{rayner10} for details) based on the Gogny \cite{gogny} EDF with the parametrization
D1M \cite{d1m}, where one notices three minima at 
spherical ($\beta\approx 0$), oblate ($\beta\approx 0.2$,
$\gamma=60^{\circ}$) and prolate ($\beta\approx 0.3$,
$\gamma=0^{\circ}$) configurations.

Intruder configurations in the IBM are introduced with the 
Duval-Barrett procedure described in Subsec.~\ref{sec:DB}, including the fact that particle-like and
hole-like bosons are not distinguished. 
The total Hamiltonian is given in Eq.~(\ref{eq:ham-cm}),
where each unperturbed Hamiltonian is specified in Eq.~(\ref{eq:ham-sg}) (plus the
$\hat L\cdot\hat L$ term), and the mixing interactions in
Eq.~(\ref{eq:mix}). 
The bosonic energy surface is now given as the obvious extension of the formula in
Eq.~(\ref{eq:pes}) to its IBM-2 version, under the assumption of 
equal proton and neutron deformation parameters.

All parameters in Eq.~(\ref{eq:ham-cm}) cannot be determined simultaneously, as
it would be a highly non-linear problem. 
Rather, they are obtained in several steps under the following  assumptions. 
\begin{itemize}

 \item[(i)] First, each unperturbed
Hamiltonian $\hat H_{n}$ ($n=0,2,4$) is fixed. 
This is done by fitting $E_{n}(\beta,\gamma)$  in each
	    diagonal matrix element of ${\cal E}(\beta,\gamma)$ in
	    Eq.~(\ref{eq:pes}) to the corresponding 
mean-field minimum by using the procedure described in Subsec.~\ref{sec:sg}. 
Based on the empirical assignment in a mean-field picture \cite{naza93}, the $0p-0h$ configuration is
	    associated with the mean-field minimum with the smallest
	    deformation (spherical minimum of the HFB total energy
	    surface in Fig.~\ref{fig:pb186-pes}). 

The same procedure is
	    applied to the $2p-2h$
configuration to describe the minimum at oblate (or second largest) 
deformation and to the $4p-4h$ configuration to describe the minimum at prolate
	    (or largest) deformation. 
In this way each unperturbed Hamiltonian $\hat H_{n}$ $(n=0,2,4)$ is
	    determined independently,
	    so as to reproduce the topology around the corresponding mean-field minimum. 

\item[(ii)] Having determined 
	    unperturbed Hamiltonians, one then calculates 
	    the energy off-set $\Delta_{2}$ ($\Delta_{4}$) so that
	    the energy difference between two neighboring
	    minima on the HFB energy surface, corresponding 
	    to the $0p-0h$ and $2p-2h$ ($2p-2h$ and $4p-4h$)
	    configurations, is reproduced. 

\item[(iii)] Finally, the mixing terms in
	     Eq.~(\ref{eq:mix}) are 
	     introduced. 
Their strengths $\omega_{n,n+2}$ ($n=0,2$) are 
	     determined so 
	     as to reproduce the overall topology of the barrier separating two neighboring mean-field
	     minima as close as possible. 
We note that the inclusion of the mixing terms affects little other
	     parameters already determined in the steps (i) and (ii): 
due to the inclusion of the mixing terms and
	     the subsequent mixing, the energy difference between 
	     neighboring minima on the energy surface changes only by a
	     few per cent in most cases. 
The above assumption appears to be valid as 
	     long as the values $\omega_{n,n+2}\approx 0.1-0.2$ MeV,
	     which are close to those used in previous configuration-mixing
	     IBM-2 calculations on Hg isotopes
	     \cite{duval81,duval82,barfield83}, are chosen. 

\end{itemize}
The parameters of the full Hamiltonian $\hat H$ in Eq.~(\ref{eq:ham-cm}) are
determined by taking these steps, and 
the resultant Hamiltonian is diagonalized in the model space
$[N]\oplus[N+2]\oplus[N+4]$. 
The corresponding mapped IBM energy surface, i.e., the lowest
eigenvalue of ${\cal E}(\beta,\gamma)$ in Eq.~(\ref{eq:pes}) at each set
of $(\beta,\gamma)$, is shown
on the right-hand side of Fig.~\ref{fig:pb186-pes}, and exhibits three
minima in accord with the HFB energy surface.

\begin{figure}[ctb!]
\begin{center}
\includegraphics[width=10cm]{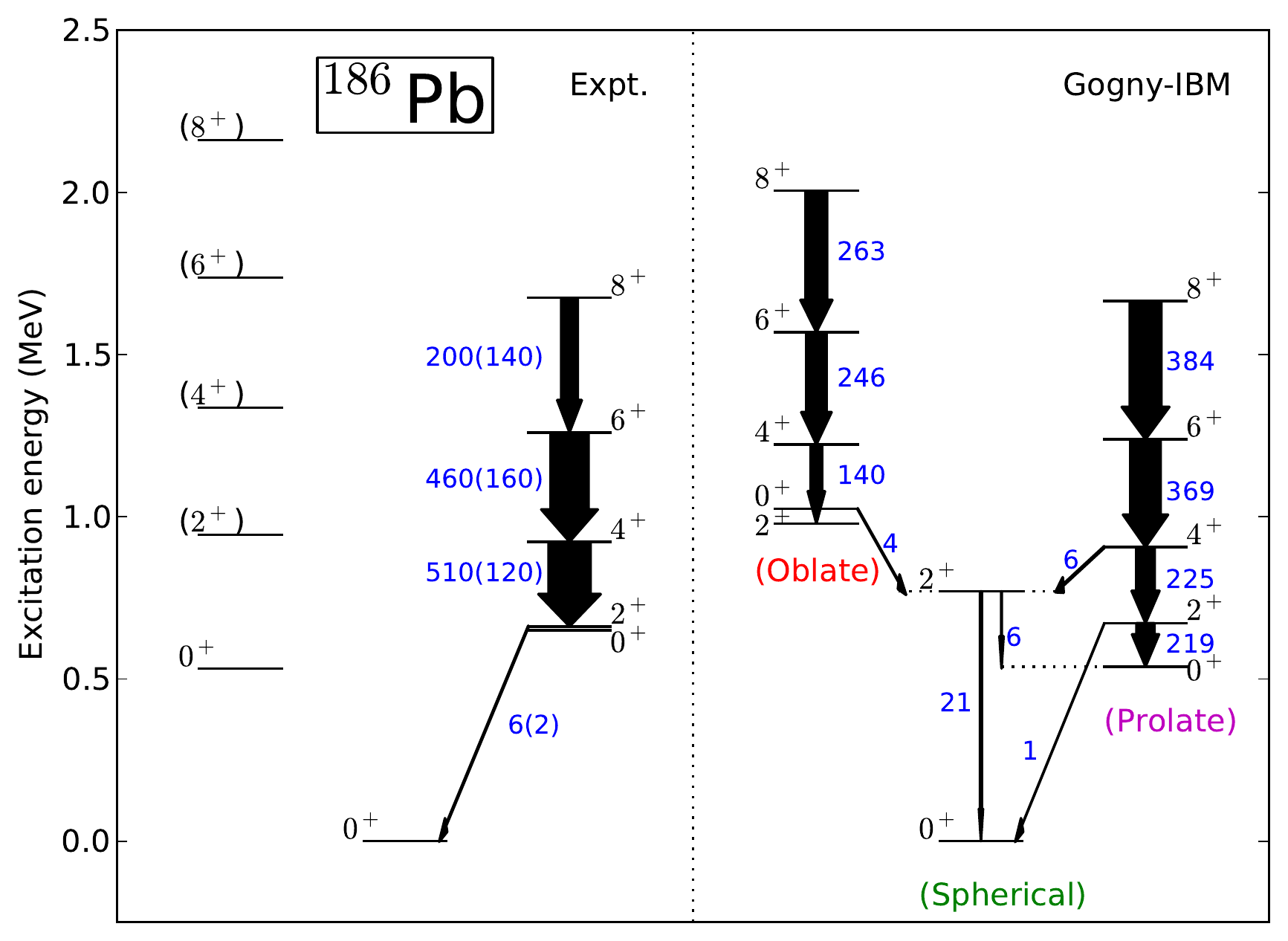}
\caption{(Color online) Experimental and theoretical (denoted by ``Gogny-IBM'') partial level schemes of
 $^{186}$Pb, including 
 two collective bands up to spin $8^+$ and the relevant $B$(E2)
 transition rates (in Weisskopf units). 
For clarity we mention that $B({\rm E2})$ values of 510(120) and
 6 (2) W.u. correspond to the $4^{+}_{1}\rightarrow 2^{+}_{1}$ and 
 $2^{+}_{1}\rightarrow 0^{+}_{1}$ transitions, respectively. 
Theoretical and experimental values are from Ref.~\cite{nom12sc} and
 Refs.~\cite{data,grahn06,pakarinen07}, respectively. }
\label{fig:pb186-level}
\end{center}
\end{figure}

In Fig.~\ref{fig:pb186-level} we compare the theoretical energy levels
with the corresponding experimental data \cite{data,grahn06,pakarinen07}. 
For the theoretical level scheme, the band assignment is made according to the $0p-0h$, $2p-2h$ and $4p-4h$
dominance of the wave function of each state and/or to the in-band E2
transition sequence. 
It is worth to note that the wave functions of low-spin states resulting from the present
IBM calculation exhibit some mixing among configurations and that this is the reason why the
labels ``Spherical'', ``Oblate'' and ``Prolate'' in
Fig.~\ref{fig:pb186-level} are put in parentheses.  
The boson effective charges for the E2 transition operator are 
taken from an earlier phenomenological IBM calculation
\cite{hellemans08}. 
Experimentally, two intruder bands are known up to spin $14^+$ and $20^+$
\cite{pakarinen07} (though with uncertainties), respectively, but  as a
benchmark 
we consider here up to spin $8^+$ states, basically for the same reason
as discussed in Subsec.~\ref{sec:proof}. 

From Fig.~\ref{fig:pb186-level}, our prediction that two quadrupole
collective intruder bands built on low-energy $0^+$ 
excited states appear and that the two excited $0^+$ states are composed of either
an oblate or prolate configuration is consistent with the experimental
finding and also agrees with the results from previous 
beyond mean-field calculations using Skyrme \cite{duguet03} and
Gogny \cite{rayner04} EDFs. 
The same level of quantitative agreement with experiment and with other
beyond mean-field calculations \cite{bender04,rayner04} 
was attained for other neutron-deficient Pb
isotopes \cite{nom12sc}.


\subsection{Systematic calculation: neutron-deficient Hg isotopes\label{sec:result}}

Now we turn to a more extensive application of the method to the Hg
isotopes, adjacent to Pb, which exhibit also the phenomenon of shape coexistence. 
The calculation in this case should include up to two configurations: regular $0p-0h$ and
intruder $2p-2h$, because as shown below the corresponding 
HFB total energy surface exhibits up to two mean-field minima.

\begin{figure*}[ctb!]
\begin{center}
\includegraphics[width=13cm]{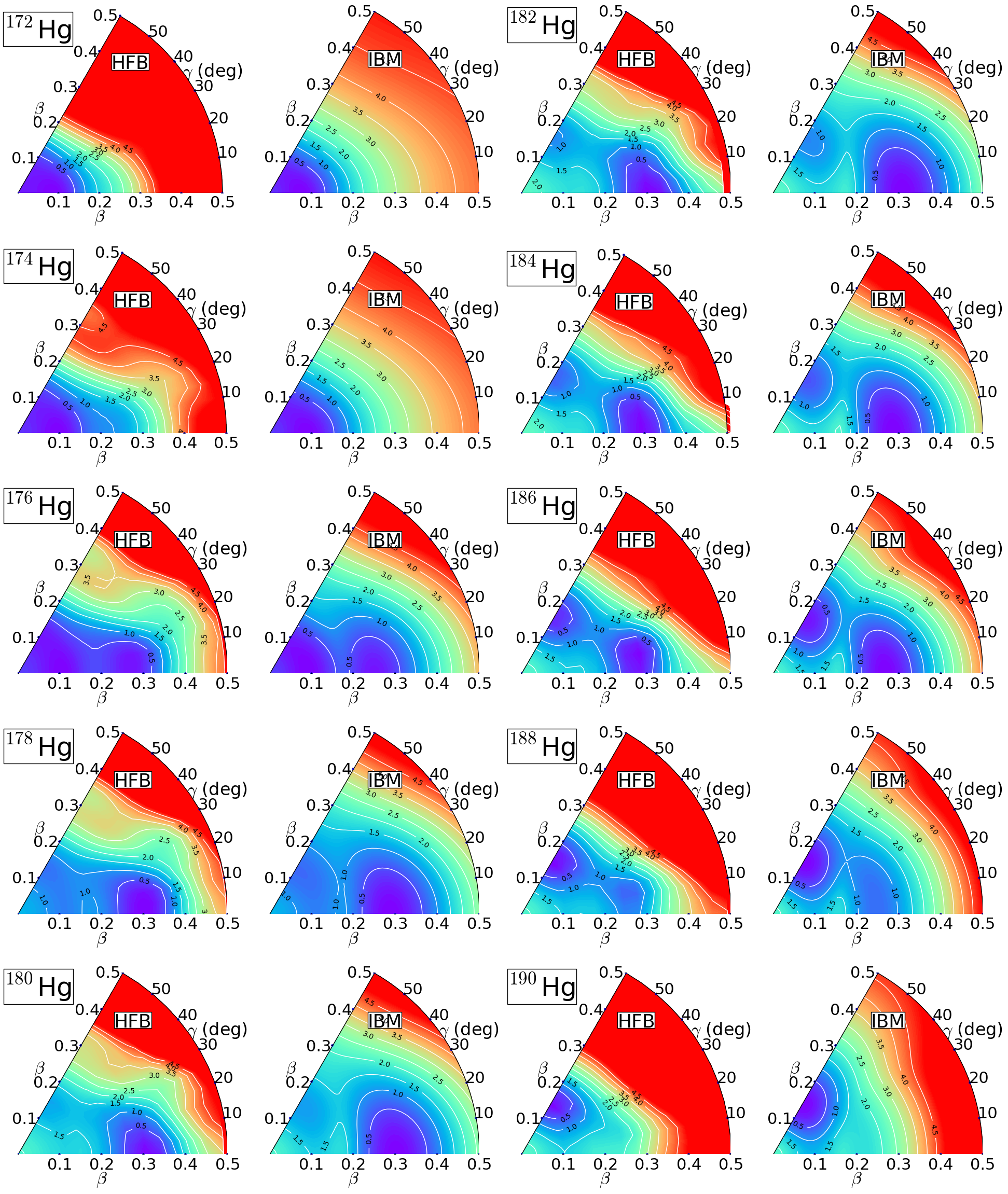}
\begin{tabular}{cccc}
\end{tabular}
\caption{(Color online) 
Same as for the caption to
 Fig.~\ref{fig:sm-pes}, but for the $^{172-190}$Hg nucleus. The results
 are taken from Ref.~\cite{nom13hg} and the HFB calculation is with the
 Gogny-D1M force. 
}
\label{fig:hg-pes}
\end{center}
\end{figure*}

In Fig.~\ref{fig:hg-pes} we display the Gogny-D1M and mapped IBM
($\beta,\gamma$) total energy surfaces of the even-even $^{172-190}$Hg isotopes. 
Starting from $^{172,174}$Hg, one observes a single minimum
corresponding to a near-spherical
configuration. 
The intruder prolate minimum appears in the $^{176}$Hg nucleus and
becomes the ground state from $^{178}$Hg to $^{184}$Hg 
corresponding to the neutron mid-shell $N=104$ 
while the second lowest minimum is on the oblate side. 
In $^{186,188}$Hg the oblate minimum becomes the ground
state and, finally, in $^{190}$Hg the prolate minimum disappears and only
a single oblate minimum is visible. 
The most pronounced prolate-oblate coexistence is observed near mid-shell. 
Both the Gogny-D1M and mapped IBM total energy surfaces give 
$\beta_2$ values consistent with previous mean-field calculations within
the Nilsson-Strutinsky method \cite{naza93,bengtsson87} and the collective
Hamiltonian approach \cite{delaroche94} based on the Gogny D1S \cite{berger84} EDF. 
Those studies predict $\beta_2\approx -0.15$ and
0.25-0.3 for the oblate and prolate configurations, respectively.

\begin{figure*}[ctb!]
\begin{center}
\includegraphics[width=13cm]{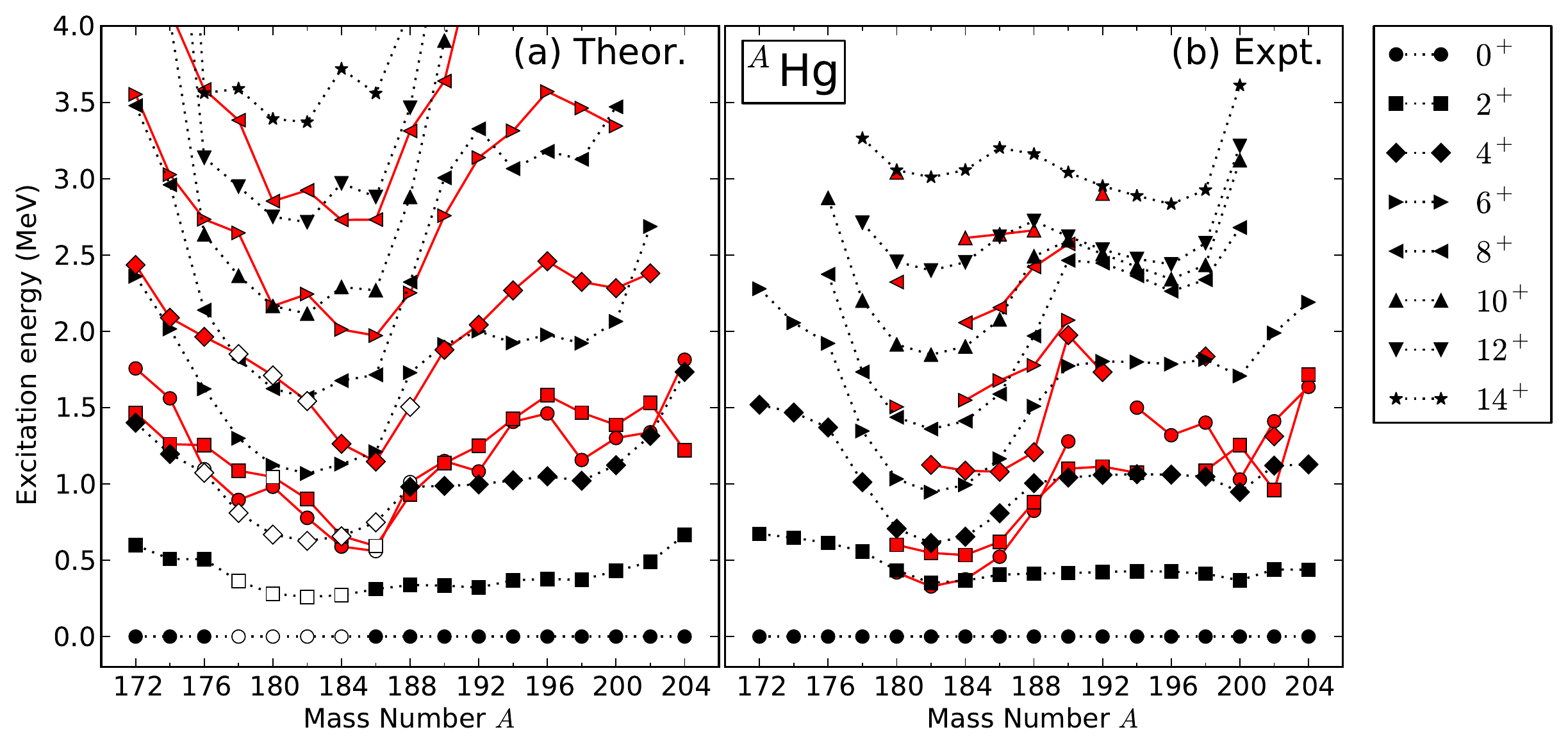}
\caption{(Color online) Theoretical (a) and experimental  
  (b) level energies of the 
 lowest two even-spin positive-parity states up to
 4 MeV for the even-even isotopes $^{172-204}$Hg. 
Dotted black and solid red lines connect the lowest and the second lowest states
 with a given spin, respectively. 
In panel (a), the $0^+$, $2^+$ and $4^+$ states, whose wave
 functions are dominated by the intruder component, are represented by open
 symbols. Theoretical and experimental values are from Ref.~\cite{nom13hg}
 and Refs.~\cite{data,sandzelius09,julin01,page11,elseviers11}, respectively. 
}
\label{fig:hg-level}
\end{center}
\end{figure*}

Next, we show in Fig.~\ref{fig:hg-level} the excitation energies of 
the lowest two even-spin positive-parity states up to spin $J=14^+$  in 
the $^{172-204}$Hg isotopes, calculated with the present method [panel (a)] 
and from the data 
\cite{data,sandzelius09,julin01,page11,elseviers11} [panel (b)]. 
Experimentally, there are many low-spin states that are not uniquely
identified and, for this reason, they are not shown in the plot. 
We first give a brief overview of the theoretical level energy
systematics 
and then compare them with the data.

Going from $^{174}$Hg to $^{176,178}$Hg, the calculated $0^{+}_2$ state
rapidly drops in energy,  
reflecting the onset of the prolate second minimum at $^{176}$Hg [see Fig.~\ref{fig:hg-pes}] and the subsequent
inclusion of the intruder configuration in the IBM calculation. 
The significant lowering of the theoretical $0^+_2$ energy level from $^{178}$Hg
toward the near mid-shell nuclei $^{184,186}$Hg 
correlates with the systematics in the HFB total energy surface showing 
coexisting minima for these nuclei. 
The $0^+$ ground states in $^{178-184}$Hg isotopes are predicted to be
of intruder nature [see
Fig.~\ref{fig:hg-level}(a)], where the corresponding total energy
surfaces predict prolate absolute minima [Fig.~\ref{fig:hg-pes}]. 
For heavier isotopes ($A\geq 186$), consistent with their
HFB total energy surfaces, the model predicts their ground states to be 
of $0p-0h$ oblate nature. 
In Fig.~\ref{fig:hg-level}(a) we observe that there is a jump from
$^{186}$Hg to $^{188}$Hg in the theoretical $0^+_2$ and other non-yrast states. 
This significant structural change is consistent with the Gogny HFB
total energy surface, where we observe that the prolate minimum is less
prominent in $^{188}$Hg than in $^{186}$Hg. 
From $^{190}$Hg onward, for which only a single oblate configuration is
required, the calculated energies of the yrast states are almost 
constant with mass number $A$, reflecting the same intrinsic 
structure; they increase when the $N=126$ shell closure is approached.

Comparing our results in Fig.~\ref{fig:hg-level}(a) with 
the data in Fig.~\ref{fig:hg-level}(b), we point out
the following features:  
\begin{enumerate}

\item First of all, the calculation describes a shape transition from 
the weakly deformed structures in the lightest Hg isotopes, to the
manifest shape coexistence near neutron mid-shell, characterized by an approximate parabolic behavior of
the non-yrast states centered around midshell, and to the weakly oblate
deformed shapes and the level structure near the $N=126$ shell
closure. 
Note, however, that the prediction on the states with higher spin
      (typically $J>8^+$) is apparently 
      not valid for the same reason as discussed in
      Subsec.~\ref{sec:proof}. 

 \item For the near mid-shell nuclei, the wave
       function content (Fig.~\ref{fig:hg-level}(a)) indicates that the lowest two 
$0^+$ states originate either from a regular $0p-0h$ or from an intruder $2p-2h$
configuration, consistent with the empirical knowledge.

\item Also for the near-midshell nuclei, particularly $^{180-184}$Hg, however, several disagreements with
      the experimental data are observed: 

\begin{enumerate}
\item  The calculation predicts the ground
       state $0^+_1$ to be intruder prolate in nature, whereas the empirical assignment is that the oblate ground
       state persists all the way and that the prolate
       intruder state emerges when approaching the neutron midshell
       $N=104$. The reason for this discrepancy is that the Gogny-D1M EDF
       total energy surface 
       predicts a prolate ground state in these Hg isotopes. 

 \item The $0^+_2$ 
       energy is substantially overestimated. In the
       present model, the
       $0^+_2$ excitation energy depends, to a large extent, on the prolate-oblate mean-field energy difference in
       the HFB energy surface. Therefore, and because of the sizable  
       level repulsion effect due to mixing, the $0^+_2$ level is pushed
       up in energy. 

 \item In the present calculation, some irregular patterns in band
       structure and $B$(E2) systematics for low-spin states are
       observed in near midshell Hg nuclei \cite{nom13hg}. 
        This arises from the
       peculiar topology of the HFB total energy surface, leading to the
       too strong mixing in low-spin states, and/or from the fact that
       the present unperturbed IBM Hamiltonians and mixing terms are too
       simple to describe the finer 
       detail of the nuclei with complex shapes. 
\end{enumerate}

\item Approaching the $N=126$ shell closure, the IBM energies of the yrast levels
      in $^{200-204}$Hg are stretched. The reason is that the IBM model space is
      too limited to describe the appearance of non-collective
      excitations when just a few neutron holes outside of $N=126$ are
      present. 

\end{enumerate}
These features, revealed in this particular example of the Hg isotopes, 
reflect some generic aspects of the present method when it is applied to the description of complex 
shape dynamics. 
These points will be detailed in the next section.

\section{Conclusions and Outlook\label{sec:summary}}

We have given a brief overview of studies of shape coexistence
and the related spectroscopy 
in the framework of the IBM. 
To describe the phenomenon of shape coexistence, the model
space including intruder excitations becomes exceedingly large in heavy
open-shell nuclei, making the 
large-scale shell-model calculation difficult. 
The IBM is an approximation exploiting effective bosonic degrees of
freedom and as such it is a drastic simplification of the nuclear
structure calculations while preserving essential ingredients of low-energy
quadrupole modes. 
This allows one to access heavy open-shell nuclei.   
Many attempts have been made in the past to incorporate intruder excitations in the IBM
framework and they have been quite successful in the description of both the spectroscopic
and ground-state 
properties which signal the shape coexistence phenomena in medium- 
and heavy-mass nuclei.

Here, special focus has been placed on the method
developed recently by linking the self-consistent HFB approach, 
or equivalently the microscopic EDF framework, and the IBM. 
Our starting point is the HFB calculation of 
the total energy surface and, by mapping it onto the energy expectation
value in the boson coherent state, the bosonic Hamiltonian is determined. 
The HFB calculation with a given EDF represents an optimal way of studying
various mean-field properties all over the periodic table, and the topology of its total
energy surface in the vicinity of the global minimum reflects features
of many-fermion systems relevant to low-energy nuclear structure. 
By establishing a mapping between fermionic and bosonic total energy
surfaces, those features can be simulated by the IBM system in a
computationally much simpler way. 
The method is general so that the IBM Hamiltonian used for the
spectroscopic study can be derived for any nuclear shape. 
This also solves the long-standing problem how to derive/understand
the IBM in a comprehensive way by starting from only fermionic degrees of freedom.

The method has opened up the possibility to describe the phenomenon of shape
coexistence in nuclei and its relevant spectroscopic properties. 
By making use of the procedure of Duval and Barrett \cite{duval81}, for the
implementation of intruder configurations in the IBM, and of Frank et 
al. \cite{frank04}, for the coherent state for 
shape coexistence, the method was extended so that it describes complex shape coexistence phenomena with a
microscopic input from EDF. 
The benchmark calculation for the neutron-deficient Pb isotopes indicates 
that the model description is sound 
in describing the two collective intruder bands built on the low-energy $0^+$ excited states associated
with oblate and prolate configurations. 
We have also shown results for neutron-deficient Hg isotopes, which are
in a reasonable agreement with existing data as regards ground- and excited-state 
properties, including the empirical finding that the low-lying
structure in the mid-shell Hg nuclei is comprised of the coexistence 
and mixing of oblate ground-state and prolate intruder configurations.

Nevertheless, the examples shown here also suggest the need for improvements and extensions of the model.

The first aspect concerns the assumptions that both the boson
Hamiltonian used and the mapping procedure
are unique, and  that the Duval-Barrett procedure to explicitly incorporate the intruder excitations in the
IBM is valid. 
Actually, as each unperturbed Hamiltonian is not of a general form,
some terms might be missing which are important to
describe correctly specific low-lying states. 
The employed unperturbed Hamiltonians and mixing interactions might be of too simplified a form to reproduce
every detail of the HFB total energy surface, which is more
complicated 
(softer) than the bosonic energy surface, and to extract all the
parameters in the unperturbed Hamiltonians, mixing strengths, off-set
energies and effective charges for transition operators. 
To constrain better the IBM Hamiltonian, it may not be sufficient to
refer to a mean-field total energy surface only, and 
the development of an advanced mapping might be required, invoking  
additional quantities, e.g., a symmetry-projected energy surface. 
Moreover, while we followed the assumption of Duval and Barrett that both particle and
hole pairs are mapped onto the same boson, a more realistic formulation would be to consider separate mappings onto
particle- and hole-like bosons and to use a single IBM Hamiltonian,
instead of introducing several different Hamiltonians.

The second aspect concerns the assumption that the EDF framework gives
the correct mean-field properties, including the total energy surface,
which is our
starting point to construct the IBM Hamiltonian. 
In fact, in some of the neutron-deficient Hg isotopes a prolate
ground-state is predicted at the mean-field
level, while it is known empirically that a weakly-oblate-deformed
vibrational band stays the ground state all
the way through the Hg chain and that the prolate
intruder state comes lowest in energy around the mid-shell $N=104$. 
This peculiar feature of the mean-field result is
naturally inherited into the mapped IBM description, and it can thus 
happen that the intruder prolate band becomes the ground-state band in some of
the Hg nuclei. 
However, it is worth to remark that the majority of currently 
available non-relativistic \cite{yao13,delaroche94,bender06} and
relativistic \cite{nik02} EDFs, whose parameters have
been determined by a global fit, commonly predict a prolate ground state at 
the mean-field level (and beyond) for some of the neutron-deficient Hg
isotopes. 
The exception is the so-called NL-SC set \cite{nik02} as it has been tailored to reproduce
the oblate ground state in the corresponding Hg isotopes. 
The problem of prolate-oblate balance in the mean field is partly attributed to the
single-particle spectrum, which determines the shell
gap 
and the energy to create intruder 
excitations of major importance. 
In that sense, as investigated in Ref.~\cite{nik02}, it would be of interest
to assess the capacity of the EDF framework to predict complex shape phenomena as
well as related spectroscopy.

Finally, the method
to implement intruder excitations (or more than one mean-field minimum) in
the IBM framework is general and could be applied to treat any deformation energy surface and
the corresponding collective dynamics. 
This method therefore paves the way to study large-amplitude collective motions like spontaneous fission. 
Such a study is certainly beyond the capability of current
computational power, especially as more degrees of freedom (e.g.,
triaxial, octupole, {\it etc}.) need to be included. 
It is nevertheless an interesting subject, and work along this line is in
progress.

\ack
We would like to thank L. Guo, T. Nik\v{s}i\'c, L. M. Robledo,
R. Rodr\'iguez-Guzm\'an, N. Shimizu, and D. Vretenar, 
for their contributions to the works reviewed in this article. 
K.N. acknowledges the support 
from the Marie Curie Actions grant within the
Seventh Framework Program of the European Commission under Grant
No. PIEF-GA-2012-327398. 

\section*{References}

\bibliography{refs}

\end{document}